# Length Matters: Keeping Atomic Wires in Check

Brian Cunningham, Tchavdar N. Todorov and Daniel Dundas
Atomistic Simulation Centre, School of Mathematics and Physics,
Queen's University Belfast, Belfast BT7 1NN, U.K.

## ABSTRACT

Dynamical effects of non-conservative forces in long, defect free atomic wires are investigated. Current flow through these wires is simulated and we find that during the initial transient, the kinetic energies of the ions are contained in a small number of phonon modes, closely clustered in frequency. These phonon modes correspond to the waterwheel modes determined from preliminary static calculations. The static calculations allow one to predict the appearance of non-conservative effects in advance of the more expensive real-time simulations. The ion kinetic energy redistributes across the band as non-conservative forces reach a steady state with electronic frictional forces. The typical ion kinetic energy is found to decrease with system length, increase with atomic mass, and its dependence on bias, mass and length is supported with a pen and paper model. This paper highlights the importance of non-conservative forces in current carrying devices and provides criteria for the design of stable atomic wires.

## INTRODUCTION

With the miniaturization of electronic devices, stability issues and failure due to large current-induced forces (CIF)—as a result of the huge current densities flowing—become a central issue. One particular component of the CIF is the electron wind force [1,2], which has received a lot of attention recently due to its non-conservative nature [1,3-5]. Non-conservative forces (NCF) could cause adverse effects in nanoscale devices. For example, NCF (as opposed to Joule heating) may be the primary mechanism behind certain electromigration phenomena and anomalous heating in atomic wires [6]. However, NCF may also be exploited constructively, for example, in the development of a nanoscale engine [7].

Long defect free atomic wires are prime candidates for observing certain properties and characteristics of NCF. The reason is the dense vibrational mode frequency spectra of these systems [8] together with their high conductivity. It is shown in [3,8] that current can couple near degenerate modes to create new modes that grow in time (referred to as waterwheel modes). In this paper we employ two methods for investigating NCF: a Landauer steady-state approach; and non-equilibrium, non-adiabatic molecular dynamics (MD) simulations. We investigate the normal modes under bias in the Landauer steady-state [9], and as we shall see, this allows us to predict the outcome of the MD. In the MD, we find that as a result of the competition between NCF and electronic friction, the typical kinetic energy of an ion (and hence effective temperature) attained under the above non-equilibrium conditions decreases with system length but increases with atomic mass. A simple model to support the findings is given in the results.

## METHOD

The types of systems investigated are illustrated in figure 1. A device region is connected to two long electrodes, which are themselves connected to electron reservoirs far from the device. The difference in the electrochemical potentials of the reservoirs creates a bias.

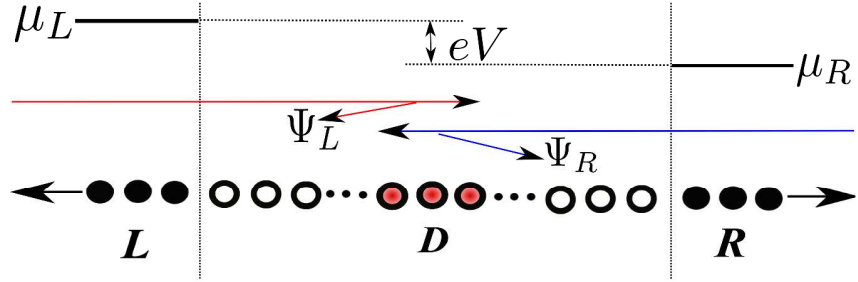

**Figure 1.** Device region *D* is connected to electrodes *L* and *R*. In the steady-state calculations (see text), the electrodes are infinite and we employ the Landauer formalism to model electron transport. In the dynamical simulations, the electrodes are long but finite and connected to external probes (see text). A subset of device atoms (red) are treated dynamically. A net flux of electrons is set up by the difference in electrochemical potentials $eV = \mu_L - \mu_R$.

The electronic structure is described in a spin degenerate 1s orthogonal tight-binding model [10] with non-interacting electrons. Tight-binding parameters for gold are chosen. The lattice spacing was set to 2.37 Å, somewhat below the equilibrium bondlength to suppress a Peierl's distortion [11] (and resultant band gap in the electronic structure) that occurs during relaxation of these systems. The hopping integral ($H_{mn} = \int \varphi^*(\mathbf{r} - \mathbf{R}_m) \hat{H}(\mathbf{r}) \varphi(\mathbf{r} - \mathbf{R}_n) d\mathbf{r}$, where $\varphi(\mathbf{r} - \mathbf{R}_{m(n)})$ is an atomic orbital centered at site *m(n)*) was −4.78 eV.

We begin by examining the steady-state properties of the system within the Landauer picture of conduction [9]. We adopt an adiabatic approach here, where—as the atoms move— quantities are assumed to not depend explicitly on time. Expectation values of electronic quantities are determined from the 1-electron steady-state density matrix,

$$\hat{\rho}(V,\vec{R}) = \int_{-\infty}^{+\infty} \left[ f_L(E) \hat{D}_L(E) + f_R(E) \hat{D}_R(E) \right] dE . \qquad (1)$$

Here *V* is bias, $\vec{R}$ are the coordinates of the classical ions, and $\hat{D}_i(E)$ and $f_i(E)$, *i=L, R*, are the partial density of states operators and occupancies for electron states arriving from electrode *i*.

The behavior of the system under small displacements about a reference geometry can be understood from the dynamical response matrix [4,8] $K_{\nu\nu'}(V,\vec{R}) = -\partial F_\nu(V,\vec{R})/\partial R_{\nu'} + \partial^2 P(\vec{R})/\partial R_{\nu'} \partial R_\nu$, where *P* is the sum of repulsive pair potentials and $F_\nu$ is the expectation value of the force operator $\hat{F}_\nu(\vec{R}) = -\partial \hat{H}(\vec{R})/\partial R_\nu$, with $\hat{H}(\vec{R})$ the electronic Hamiltonian. $K(V,\vec{R})$ is non-Hermitian under current [8], and has anti-symmetric part

$$A_{\nu\nu'}(V,\vec{R}) = 2\pi \sum_{i=L,R} \int_\mu^{\mu_i} \text{Im Tr}\{\hat{F}_\nu \hat{D}(E) \hat{F}_{\nu'} \hat{D}_i(E)\} dE, \qquad (2)$$

where $\mu$ is the equilibrium chemical potential and $\hat{D}(E)$ is the total density of states. The anti-symmetric part can generate non-equilibrium normal modes with complex frequencies that come in conjugate pairs and describe motion that grows/decays in time.

We also investigate the dynamical effects of CIF and the competition between NCF and electronic friction through real-time MD simulations. In these simulations current is generated by the open boundary method of [12] with 250 atoms in the electrodes and a device region of 300 atoms (a subset of which are treated dynamically). The simulations employ the Ehrenfest approximation, in which classical ions interact with the mean instantaneous electron density. The Ehrenfest approximation suppresses Joule heating, allowing us to isolate and study the influence of NCF on the ions in these chains. The additional cooling effect of lattice conduction into the electrodes is also neglected, giving us an upper bound on the effects of work done by NCF. Conduction out of the mobile region is investigated in [13].

## RESULTS & DISCUSSION

### Preliminary static calculations

We calculate the equilibrium mode frequencies for the systems in figure 1 after a geometry relaxation on the device atoms (nearby geometries produced similar phonon structure). Figure 2(a) shows the range of equilibrium frequencies as a function of the number of mobile atoms, $N$, for atoms with mass 10 a.m.u..

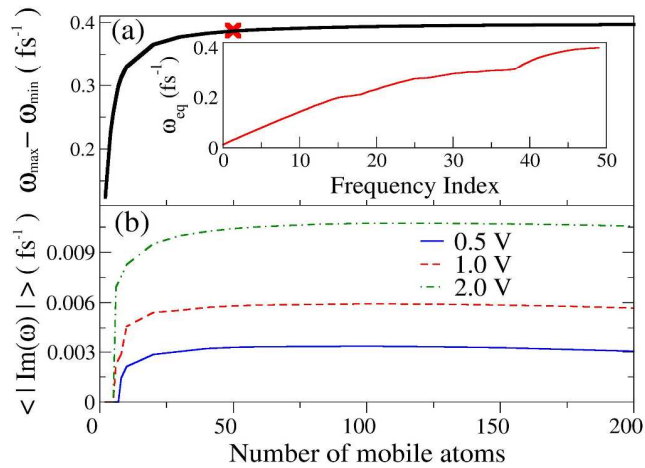

**Figure 2.** (a): Range of equilibrium eigenfrequencies as a function of the number of mobile atoms. The inset displays all frequencies for the length marked on the main curve with an **X**, $N = 50$. (b): $\Phi$ (see equation 3) as a function of $N$ for the three biases indicated.

Since frequency bandwidth saturates with $N$, we expect more near degenerate modes to couple under current forming waterwheel pairs as the mobile region length is increased. We then determine the normal modes under bias and examine their imaginary parts. The quantity

$$\Phi = \frac{1}{N}\sum_{\alpha=1}^{N} |\operatorname{Im}(\omega_\alpha)|, \qquad (3)$$

as a function of *N* for a few biases is shown in figure 2(b) and gives a measure of the average rate of work, per atom, due to NCF. The complex modes contributing to Φ are in fact a small fraction (which increases with bias) of the total number of degrees of freedom, and have closely clustered real and imaginary parts. From figure 2(b), we expect certain bias-dependent length-independent characteristics and this will now be investigated through dynamical simulations.

## **Dynamical Simulations**

The dynamical simulations start from the relaxed geometry and figure 3 shows the total ion kinetic energy as a function of time for 200 mobile atoms under 0.5 V. The kinetic energy rises initially and then the electronic friction comes in to play to eventually balance the NCF. Beyond about 3 ps, a steady-state is reached and the energy settles to a value of 17.5 ± 1.5 eV.

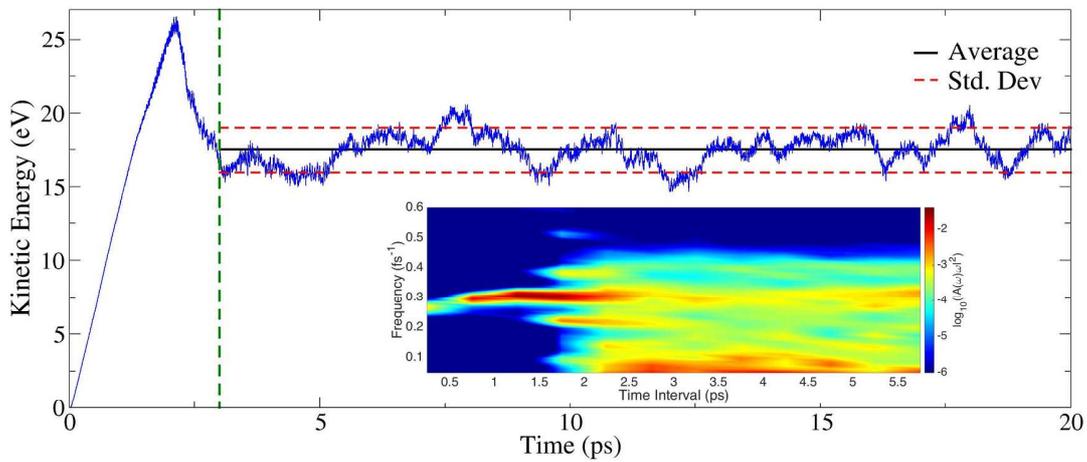

**Figure 3.** Total ionic kinetic energy versus time for 200 mobile atoms with mass 10 a.m.u., under a bias of 0.5 V. The kinetic energy increases sharply under the NCF. Due to the counter-balancing electronic friction, a steady-state is eventually reached and this allows us to determine average values for quantities such as the ion kinetic energy and electron current. The inset displays a log plot of the distribution of total ion kinetic energy (arb. units) among frequencies (obtained from a short-time Fourier transform) during the first 6 ps of the simulation.

The energy distribution among phonon frequencies (inset in figure 3) shows a prominent contribution at around 0.3 fs$^{-1}$, especially at the start of the simulation. This frequency is close to: the plateau frequency in the inset of figure 2(a); the typical real part of complex mode frequencies; and the Einstein frequency for this system, $\omega_{\text{Ein}}$. The energy is initially contained in a few modes in that range, and eventually fills the frequency band as the steady-state is attained.

Once the steady-state is reached, the NCF—proportional to current/bias (see figure 2(b)) and displacement (for small displacements)—balance the frictional forces (proportional to velocity), producing the relation

$$I \propto \frac{1}{\sqrt{M}}, \tag{4}$$

where *I* is the saturation current (time and spatially averaged steady-state bond current [8]) and $M \propto 1/\omega_{Ein}^2$. Equation 4 is tested in figure 4(a); figure 4(b) strengthens the prediction of a bias- and length-independent saturation current (for $N \gtrsim 40$ mobile atoms).

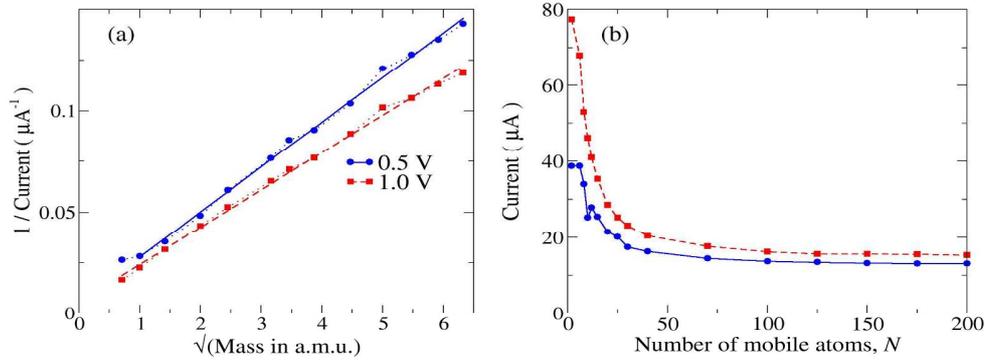

**Figure 4.** (a): one over saturation current (see text) versus square root of ion mass for a mobile region with 200 atoms. (b): Saturation current as a function of *N* for the two biases in (a) with a mass of 10 a.m.u. (Notice that for small *N*, the conductance is almost equal to one quantum unit).

Figure 5 then shows the saturation kinetic energy as a function of *N*. The inset suggests the relation *1/E = a(M,V)1/N + b(M,V)*, where *a* and *b* are the bias- and mass-dependent slope and intercept of the lines of data points.

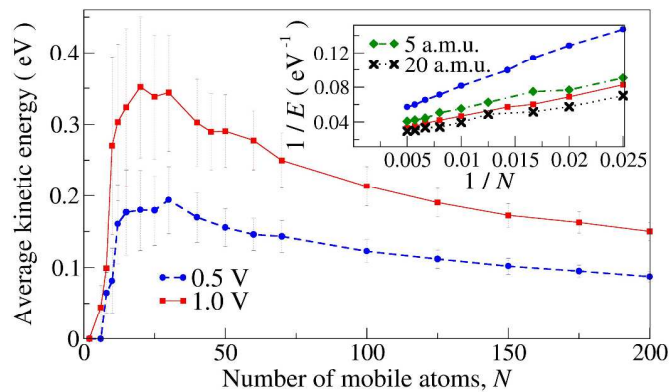

**Figure 5.** Saturation kinetic energy per atom as a function of the number of mobile atoms, *N*, for atomic mass 10 a.m.u. under the two biases stated. The inset displays one over the total energy, *E*, versus *1/N* for the two biases in the main figure, and for the 1 V case with masses 5 and 20 a.m.u.. We see a clear linear relation between *1/E* and *1/N*.

To interpret the results in figures 4 and 5 we start with the current-voltage relation for diffusive conduction [14], $I = gV/(1 + L/l)$, where *g* is the quantum conductance unit, *L* is the mobile region length and *l* is of the order of the electronic mean free path. Using equation 4 and assuming $1/l$ to be proportional to the mean energy per atom, we obtain

$$\frac{\alpha eV}{1+\frac{E}{E_0}\left(1+\frac{\beta}{N\hbar\omega}\right)} = \hbar\omega, \quad (5)$$

where $E_0$ and $\beta$ are constants. The correction in brackets in equation 5 gives agreement with the results in figure 5, and can be understood by considering the interplay between normal electron diffusion and thermally assisted hopping between localized electron states [8].

**CONCLUSIONS**

The results from the simulations are incorporated into a simple relation, equation 5, which indicates that longer wires with light atoms will be more stable and less prone to failure due to non-conservative energy transfer from the current into the atomic motion. The methods and results above provide a basis for investigating NCF in more complex and experimentally accessible systems. Paths for further work include: dynamics in the presence of the band gap that arises due to the Peirel's instability (occurring during geometry relaxation); allowing for electron-ion correlations, enabling us to investigate the interplay between NCF and Joule heating; and the effect of lattice heat conduction out of the system, as in reference [13].

**ACKNOWLEDGMENTS**

We are grateful for support from the UK **EPSRC** under grant EP/I00713X/1. This work used the **ARCHER** UK National Supercomputing Service. We thank Mads Brandbyge, Per Hedegård, Jing-Tao Lü and Jan van Ruitenbeek for fruitful discussions.